\documentclass[12pt]{article}
\usepackage{latexsym,amsfonts,amssymb,amsmath}
\usepackage[utf8]{inputenc}

\numberwithin{equation}{section}

\newcommand{\beq}{\begin{equation}} \newcommand{\eeq}{\end{equation}}
\newcommand{\bea}{\begin{array}} \newcommand{\eea}{\end{array}}
\newcommand{\ri}{{\mathrm i}}

\catcode`@=11 \long
\def\@caption#1[#2]#3{\par\addcontentsline{\csname
ext@#1\endcsname}{#1} {\protect\numberline{\csname
the#1\endcsname}{\ignorespaces #2}} \begingroup \small
\@parboxrestore \@makecaption{\csname fnum@#1\endcsname}
{\ignorespaces #3}\par \endgroup} \catcode`@=12

\newcommand{\la}{\label}

\newcommand{\so}{\sigma_1}
\newcommand{\sd}{\sigma_2}
\newcommand{\st}{\sigma_3}

\newcommand{\p}{\partial}

\begin{document}
\allowdisplaybreaks

\allowdisplaybreaks
 \begin{titlepage} \vskip 2cm

\begin{center} {\Large\bf Symmetries  and Supersymmetries of Generalized  Schr\"odinger equations }
\footnote{E-mail: {\tt nikitin@imath.kiev.ua} }\footnote{Dedicated to V\`{e}ronique Hussin
 } \vskip 3cm {\bf {A.
G. Nikitin } \vskip 5pt {\sl Institute of Mathematics, National
Academy of Sciences of Ukraine,\\ 3 Tereshchenkivs'ka Street,
Kyiv-4, Ukraine, 01004\\}}\end{center} \vskip .5cm \rm

\begin{abstract}
 The contemporary results concerning supersymmetries in generalized Schr\"odinger equations are presented. Namely, position dependent mass Sch\"odinger equations are discussed as well as the equations with matrix potentials. An extended number of realistic quantum mechanical problems admitting extended supersymmetries is described, an extended class of matrix potentials is classified.
\end{abstract}
\end{titlepage}
\section{Introduction\label{int}}\label{intro}

In seventieth of the previous century   a new qualitatively new symmetry  in
physics has been  discovered and called {\it supersymmetry} (SUSY) see, e.g., \cite{akula}  but also \cite{lipa}
where the idea of SUSY was formulated in somewhat rudimentary form. Its rather specific property is the existence of symmetry transformations mixing
bosonic and fermionic states  In other words  transformations
 which connect fields with different statistics have been introduced.

 Among the many attractive features of SUSY is that it provides an
effective  mechanism for the cancelation of the ultraviolet
divergences in quantum field theory. In addition, it opens new ways   to unify space–time
symmetries (i.e., relativistic  invariance) with internal symmetries and to construct   unified field theories, including all  types
of interactions, refer, e.g., to \cite{kos}, \cite{Green} and \cite{Sei}.

Mathematically, SUSY requests using of the graded Lie algebras instead of the usual ones, and the corresponding  group parameters are not numbers but Grasmanian variables. The essential progress in the related fields of mathematics was induced exactly by the needs of SUSY.

Unfortunately, till now we do not have convincing experimental arguments for introducing SUSY as a universal symmetry  realized in Nature. But there exist an extended number of realistic physical systems which do be supersymmetric. Moreover, SUSY presents effective tools for for understanding the relations between spectra of different Hamiltonians as
well as for explaining degeneracy of their spectra, for constructing exactly
or quasi-exactly solvable systems, for justifying formulations of initial and
boundary problems, etc., etc.; see, e.g., surveys \cite{gen,Coop,Fer} and monograps \cite{Jun,Bag}. In other words, SUSY is realized in Nature at least in a rather extended number of realistic physical systems.

The present work is concentrated on quantum mechanical systems since
they provide a ground for testing the principal question: whether SUSY is
realized in Nature or not, free of the complexities of field theories. Examples
of such systems (like interaction of spin 1/2 particle with the Coulomb or
constant and homogeneous magnetic field) which admit exact N = 2 SUSY
are well known \cite{Suka}, \cite{Rava} (see also Refs. \cite{gen}, \cite{Coop} and references therein). However, we will be concentrated on systems admitting more extended SUSY.

Let us remain that the supersymmetric quantum mechanics was created by Witten \cite{Witten} as a toy model for illustration of global properties of the  quantum field theory. But rather quickly it becomes a  fundamental
field attracting the interest of numerous physicists and mathematicians.
In particular the SSQM  presents powerful tools for explicit
solution of quantum mechanical problems using the shape invariance
approach \cite{Gen}. The number of problems
satisfying the shape invariance condition is rather restricted but includes the majority of exactly solvable
Schr\"odinger equations. The well known exceptions are exactly
solvable Schr\"odinger equations with  Natanzon potentials
\cite{natan} which are formulated in terms of implicit functions.

A very important application of SUSY in quantum mechanics is classification of families of isospectral Hamiltonians. And there is a number of systems isospectral with the basic exactly solvable SEs. In the standard SUSY approach with the first order intertwining operators the problem of description of such families is reduced to constructing general solutions of the Riccati equations. More refined approaches can include intertwining operators of higher order \cite{andr}, the N-fold supersymmetry \cite{tana}  and the hidden nonlinear
supersymmetry \cite{Pliuha}. One more relevant subject of contemporary SUSY  are so-called exceptional orthogonal
polynomials \cite{opa2}, \cite{buba}.

Let us mention that other generalized supersymmetries which include the usual SUSY have been discussed also, among them the so called parasupersymmetry \cite{Ruba,Nata,bekki1,be1}, which  has good roots in real physical problems \cite{bebe}. However, the standard SUSY is seemed to be more fundamental.

Just in quantum mechanics SUSY presents powerful tools
for constructing exact solutions of Schr\"odinger equation (SE). And we will present a survey of contemporary results belonging to this fields. We will not discuss   generalizations of the standard SUSY in quantum mechanics like the ones  mentioned above,  but restrict ourselves to the standard SUSY quantum mechanics with the first order intertwining operators \cite{bek1}. However, the systems with extended SUSY as well as systems including SEs with Pauli and spin-orbit couplings,  with position dependent mass and  with abstract matrix potentials will be considered. Notice that just these fields are the subjects of current interest of numerous investigators.

Let us stress that there are two faces of SUSY in quantum mechanics. First, there exist QM systems like the charged particle with spin 1/2 in the constant and homogeneous magnetic field which admit exact SUSY.  Such systems admit constants of motion forming superalgebras.  Secondly, it is possible to indicate the QM systems with "hidden" SUSY like the Hydrogen atom,  and just these systems can be solved exactly using the shape invariance of the related Schr\"odinger equations. We will discuss both types of SUSY. The
 realistic physical systems which admit exact SUSY will be be considered in the next section, while the shape invariant systems are discussed in Sections 3-6.

 An inspiring example of QM problem with a shape invariant potential was discovered by Pron’ko and Stroganov
\cite{Pron} who studied a motion of a neutral non-relativistic
fermion, e.g., neutron, interacting with the magnetic field
generated by a  current carrying wire. A relativistic version of such problem was discussed in \cite{ninni}.

 The specificity
 of the PS problem is that it includes  a {\it matrix superpotential} while in the standard SUSY in quantum mechanics
  the superpotential is a scalar function. Matrix potentials and
  superpotentials naturally appears  in  quantum mechanical models
  including
  particles with spin (see, e.g., \cite{Khare}, sections 10 and 11) and  in multidimensional models of SSQM
  \cite{iof1,iof2}. Particular examples of such superpotentials was discussed in
  \cite{Andr,Andri,Fu,Ioffe,Rodr}. In papers \cite{tkach}
  such superpotentials were used for modelling the motion of a spin
  $\frac12$ particle in superposed magnetic and scalar fields.
  In paper \cite{Fu} a certain class of such superpotentials was described, while more extended classes of them were classified in \cite{138}, \cite{139}.
  In any case just  matrix superpotentials belong to an interesting research field which makes   it
  possible to find new coupled systems of exactly solvable Schr\"odinger equations. The contemporary results in this field will be discussed in the following.

  In addition to SUSY, some SEs can posses one more nice property called superintegrability (SI). By definition, the quantum system is called superintegrable if it admits more integrals of motion than the degrees of freedom. Like  SUSY, the SI can cause the exact solvability of the related SE, especially in the case when it is the maximal SI when the number of integrals of motion is equal to $2n+1$ where $n$ is the number of the system degrees of freedom.

  As a rule superintegrable systems admit higher order integrals of motion realized by differential operators of order higher than one and even higher than two \cite{anik,wintwint,snobl}. Such integrals of motion have various interesting applications including the construction of non-standard conservation laws \cite{70}.

  There exist a tight connection between the SI and SUSY, and many QM systems are both supersymmetric and superintegrable. In fact the maximal SI induces SUSY and vice versa, in spite of that this fact was never proven for generic QM systems.

  The superintegrable systems which are also supersymmetric will be a special subjects of our discussion. Moreover, they will be the systems with position dependent masses which are discussed in Section 6.

\section{QM systems with exact SUSY}
\label{sec:1}

\subsection{System with $N=2$ SUSY}
\label{sec:2}

%Don't forget to give each section
%and subsection a unique label (see Sect.~\ref{sec:1}).

Let us start with the well known and important physical system, i.e., the spinning and charged particle interacting with an external  magnetic field. The corresponding QM Hamiltonian can be written in the following form:
 \begin{equation}\la{1}H=\frac{\pi^2}{2m}+\frac{ e}{2m}{ \sigma_i}{ B_i}\end{equation}
 where
 $\pi^2=\pi_1^2+\pi_2^2+\pi_3^2,\ \ \pi_i=-\ri\frac\p{\p x_i}-eA_i,\ \ i=1,2,3,\ B_i=\varepsilon_{ijk}\frac{\p A_j}{\p k},\ \ \sigma_i$ are Pauli matrices, $B_i$ and $A_i$ are components of the external magnetic field and the corresponding vector-potential, $\varepsilon_{ijk}$ is the totally antisymmetric unit tensor, and summation is imposed over the repeating index $i$.

 In contrast with the standard Schr\"odinger Hamiltonian, operator (\ref{1}) includes the Pauli term $\frac{ e}{2m}{ \sigma_i}{ B_i}$ describing the interaction of the particle spin with the external magnetic field. The related stationary Schr\"odinger equation has the standard form
 \begin{equation}\la{00}H\psi=E\psi\end{equation}
 with $E$ being the Hamiltonian eigenvalues.

 In the case of the constant and homogeneous magnetic field directed along the third coordinate axis  the vector-potential can be reduced to the form:
 \begin{equation}\la{0} A_1=-\frac12 x_2B_3,\quad A_2=\frac12 x_1B_3, \quad A_3=0 \quad \end{equation}
  and by definition $B_1=B_2=0, B_3=B=const.$ Thus Hamiltonian (\ref{1}) can be rewritten in the following form:
 \begin{equation}\la{2}H=H_1+H_2, \quad H_1=\frac{p_3^2}{2m},\quad H_2=\frac{(\sigma_1\pi_1+\sigma_2\pi_2)^2 }{2m}\end{equation}
with $p_3=-\ri\frac\p{\p x_3}$.

The immediate consequence of representation (\ref{2}) is that our Hamiltonian commutes with the three  operators:
\begin{equation}\la{3}Q_1=\sigma_1\pi_1+\sigma_2\pi_2, \ Q_2= \ri\sigma_3 Q_1, \ Q_3=p_3\end{equation} which satisfy the following algebraic relations:
\begin{equation}\la{4}[Q_3,Q_1]=[Q_3,Q_2]=[Q_3,H]=0,\end{equation}\begin{equation}
\la{5}
\{Q_\mu,Q_\nu\}=2\delta_{\mu\nu}H_2, \ [Q_\mu,H_2]=0\end{equation}
where $\mu, \nu$ independently takes the values $1, 2$, $\delta_{\mu\nu}$ is the Kronecker delta, the symbols [.,.] and \{.,.\} denote the commutator and anticommutator correspondingly.

Less evident but also well known are the following constants of motion which commute with Hamiltonian (\ref{1}):
Less evident but also well known are the following constants of motion which commute with Hamiltonian (\ref{1}):
\begin{gather}\la{lj}\begin{split}&J_3=x_1p_2-x_2p_1+\frac12\sigma_1,\\&
K_1=\p_1-\frac12x_2B_3,\quad K_2=\p_2+\frac12x_1B_3 \end{split}\end{gather}
where $p_1=-\ri\frac{\p}{\p x_1}$ and $p_2=-\ri\frac{\p}{\p x_2}$.

Operator $J_3$ is the third component of the total angular momentum while $K_1$ and $K_2$ are Johnson-Lippmann constants of motion. They are rather similar to our $\pi_1$ and $\pi_2$ but have the opposite  sign  for $A_1$ and $A_2$.

 Thus the considered Hamiltonian admits six constants of motion. Four of them,  , i.e., $P_3, J_3, K_1$ and $K_2$  commute between themselves and with $Q_1$ and $Q_2$
 . On the other hand, $Q_1$ and $Q_2$ are not in involution, but satisfy more complicated relations (\ref{5}), which characterize a {\it Lie superalgebra}.

  Just this specific supersymmetry can be treated as the reason of the two fold degeneration of the Landau levels, i.e. the non-ground energy levels of a spin 1/2 particle interacting with the constant and homogeneous magnetic field.

 Generally speaking,  superalgebra is a graded algebra. In the simplest case of the $Z_2$ grading the elements of the superalgebra belong to two different classes, say, are odd or even. The multiplication lows for even and odd elements are different. In our case $Q_1$ and $Q_2$ are odd while $Q_3$, $H_1$ and $H_2$ are even. The product of two algebra  elements is defined as the commutator if at least one of them is even and as the anticommutator if both the elements are odd. In SUSY quantum mechanics the odd elements are called superchages. Since we have indicated two supercharges then it is possible to indicate the $N=2$ SUSY admitted by the considered system.

\subsection{Extended SUSY }

 The considered system is only a particular (albeit very important) example of realistic physical problem admitting exact supersymmetry. In particular, it is obvious that the presented SUSY is valid for arbitrary Hamiltonian admitting representation (\ref{1}) provided one component of the vector-potential of the external field is identically zero.

We will discuss also another examples, but first let us note that in fact equation (\ref{00}) with Hamiltonian (\ref{2}) admits a more extended SUSY.

 In analogy with the above we can construct a supercharge valid for equation (\ref{1}) in the case of arbitrary external magnetic field:      \begin{equation}\la{6} \tilde Q_1=\sigma_i\pi_i\end{equation}
  since $\tilde Q_1^2= H$.

  Let us show that it is possible to find three more supercharges
provided the external field is given by relations (\ref{0}). To do it we exploit the fact that equation (\ref{2})
 is invariant w.r.t. the following three discrete
transformations:
\begin{equation}\la{7}\psi\to R_3 \psi,\quad \quad \psi\to CR_1 \psi,\quad \quad \psi\to CR_2 \psi,\end{equation}
where $R_a\ (a=1,2,3)$ are the space reflection transformations
\begin{equation}\la{8}R_a \psi({\bf x})=\sigma_a \theta_a \psi({\bf x}),\quad \quad \theta_a \psi({\bf
x})=\psi(r_a {\bf x}).\end{equation}
Here
\begin{equation}\la{9}r_1 {\bf x}=(-x_1,x_2,x_3),  \quad
r_2 {\bf x}=(x_1,-x_2,x_3),\quad
r_3 {\bf x}= (x_1,x_2,-x_3),\end{equation}
and $C=i\sigma_2c$, where $c$ is the operator of complex conjugation
\begin{equation}\la{10}c \psi({\bf x})= \psi^*({\bf x}).\end{equation}

Note  that  operators ({\ref{7}) satisfy the following relations
\begin{equation}\la{10a}\begin{array}{l}\left \{ R_a,\sigma_i \pi_i  \right \} = \left \{ R_a,C
\right \} = \left \{ CR_1,\sigma_i\pi_i  \right \} = \left
\{ CR_2,\sigma_i \pi_i  \right \} =0, \\
R_a^2=-C^2=1,\quad \quad a=1,2,3.\end{array}\end{equation}
     Using (\ref{6} ), (\ref{10a}) we can see that the operators
\begin{equation}\la{11}\tilde Q_1=\sigma_i \pi_i\quad  Q_2=iR_3 \tilde Q_1 ,\quad
 Q_3=CR_1\tilde Q_1,\quad  Q_4=CR_2 \tilde Q_1\end{equation}
fulfill  the following relations
\begin{equation}\la{12}\left \{ Q_k,Q_l \right \}  =2g_{kl} \hat H, \quad \quad \left[Q_k,\hat H
\right]=0,\end{equation}
where $k,l=1,2,3,4, g_{11}=g_{22}=-g_{33}=-g_{44}=1;\  g_{kl}=0,
k\neq l.$ In other words, operators (\ref{11}) are supercharges
generating the $N=4$ extended SUSY.

Let us note that the main trick  for constructing the extended SUSY was using the discrete involutive symmetries, i.e., reflections (\ref{8}), (\ref{9}). We will see  that in analogous way it is possible to find extended SUSY for rather generic equations (\ref{00}).
\subsection{Extended SUSY with arbitrary vector-potentials}
The results of the previous section can be generalized  to extended class of arbitrary potentials with well defined parities.
       Starting with reflections (\ref{8}) we find that the
corresponding parity properties of vector-function {\bf A}({\bf x})
(\ref{0}) are of  the form
\begin{equation}\la{par}{\bf A}(r_1 {\bf x})=-r_1 {\bf A}({\bf x}),\quad  {\bf A}(r_2 {\bf x})=-r_2
{\bf A}({\bf x}),\quad  {\bf A}(r_3 {\bf x})=r_3 {\bf A}({\bf x}). \end{equation}

       Relations (\ref{par}) are satisfied by a large class of potentials
which includes (\ref{0}) as a particular case. For all such potentials the corresponding
equation (\ref{00}) is invariant w.r.t. involutions (\ref{7}) and so
admits the extended SUSY generated by supercharges (\ref{11}).
Moreover, equation (\ref{1}) for $g=2$ and an arbitrary uniform
magnetic field , i.e., the field
 \begin{equation}A_1=A_1(x_1,x_2),\quad  A_2=A_2(x_1,x_2),\quad  A_3=0, \end{equation}
admits all internal symmetries described in previous section
provided {\bf A}({\bf x}) satisfies relations (\ref{par}).

       Other systems with extended SUSY can be found by extending
reflections (\ref{9}) to the eight-dimensional group of involutions,
i.e., by adding the  fixed {\it rotation} transformations
\begin{equation}\la{15}\begin{array}{c}r_{12} {\bf x}=(-x_1,-x_2,x_3),\quad  r_{31} {\bf x}=(-x_1,x_2,-x_3),\\ r_{23}
{\bf x}=(x_1,-x_2,-x_3), \quad
 r_{123} {\bf x}=(-x_1,-x_2,-x_3).\end{array}\end{equation}

        Let the vector potential ${\bf A}({\bf x})$ has definite
parities w.r.t. a subset of transformations (\ref{9}) and (\ref{15}). Than it is possible to construct supercharges which generate extended $N=4$ and even $N=5$ SUSY \cite{Nied1}.

Thus we present a number of SE admitting extended SUSY. Let us stress then among them there is a lot of systems with a clear exact physical meaning, see \cite{Nik1}   for discussion of this aspect.

\section{SUSY in one dimension and shape invariance}
The models considered in the above are two or three dimensional in spatial variables and include systems of coupled Shcr\"odinger equations. However, many of them can be reduced to one dimensional systems using the separation of variables. Moreover, these systems can be decoupled.

Returning to equation (\ref{00}) for a charged particle interacting with the constant and homogeneous magnetic field we can exploit its rotational invariance and search for solutions in separated radial and angular variables, i.e., to represent the wave function $\psi$ as
\begin{equation}\la{n1}\psi=\frac1{\tilde r}R(\tilde r)\mathrm{e}^{n\varphi}\end{equation}
where $\tilde r=\sqrt{x_1^2+x_2^2},\ \varphi=\arctan\frac{x_2}{x_1}$. As a result we come to the following equation for the radial functions:
\begin{equation}\la{n2}\tilde H R\equiv \left(-\frac{\partial^2}{\partial \tilde r^2}-\frac{m(m+1)}{\tilde r^2} +\omega \sigma_3+\omega^2\tilde r^2\right)R=\tilde E R\end{equation}
where $m=n-\frac12$, $\omega=2m\alpha$ and $\tilde E=2mE+q_3^2+\omega n$.

Alternatively, using the gauge transformation it is possible to pass from vector-potential (\ref{0}) to the following ones: $A_1=eHx_2,\quad  A_2=A_3=0.$ Then, representing the wave function in the form $\psi=\exp[\ri (p_1x_1+p_3x_3)]\phi(x_2)$ and setting $x_2=\frac1{\alpha B}(p_1+\sqrt{\alpha B}y)$ we obtain the following equation for $\phi$:
\begin{equation}\la{n3}\hat H\phi=\hat E\phi,\end{equation}
where
\begin{equation}\la{n4}\hat H=-\frac{\p^2}{\p y^2}+\sigma_3\omega+\omega^2x^2,\quad \hat E=2mE-p_3^2.\end{equation}

Equation (\ref{n4}) defines the supersymmetric oscillator while (\ref{n2}) is rather similar to the "3d supersymmetric oscillator" but includes half integer parameter $m$ while in the $3d$ oscillator this parameter is integer. Both the mentioned equations are decoupled to direct sums of equations since the related Hamiltonians $\tilde H$ and $\hat H$ have the following form:
\begin{equation}\la{n5}\tilde H=\left(\begin{array}{ll}\tilde H_+&0\\0&\tilde H_-\end{array}\right),\qquad
\hat H=\left(\begin{array}{ll} \hat H_+&0\\ 0&\hat H_- \end{array}\right)\end{equation}
where
\begin{equation}\la{n7}\tilde H_\pm=-\frac{\p^2}{\p\tilde r^2}+\frac{n(n\mp1)}{\tilde r^2}+\omega^2\tilde r^2\pm\omega,\quad \hat H_\pm=-\frac{\p^2}{\p\tilde y^2}+\omega^2\tilde y^2\pm\omega.\end{equation}
Hamiltonians $\hat H_\pm$ have two nice properties. First, they can be factorized:
\begin{equation}\la{n8}\hat H_+=a^+a,\quad \hat H_-=aa^+\end{equation}
where $a^+$ and $a^-$ are the first order differential operators:
\[a^+=-\frac\p{\p y}+W,\quad a^-=\frac\p{\p y}+W\]
with $W=\omega y$. Secondly, these Hamiltonians coincide up to a constant term:
$\hat H_+ =\hat H_-+2\omega.$

Hamiltonians $\tilde H_\pm$ are factorizable too:
\begin{equation}
\label{n9}
\tilde{H}_-=a_\kappa^+a_\kappa^-+c_\kappa, \quad \tilde{H}_+=a_\kappa^-a_\kappa^++c_{\kappa+1}
\end{equation}
where
\begin{equation}\la{n11}
a_\kappa^-=\frac{\partial}{\partial x}+W_\kappa,\quad a_\kappa^+=-
\frac{\partial}{\partial x}+W_\kappa,
\end{equation}
and $
c_\kappa={(2\kappa-1)\omega}.$ Moreover these Hamiltonians satisfy the following relation
\begin{equation}\la{n13}\tilde H_+(\kappa)=\tilde H_-(\kappa+1)+C_\kappa\end{equation}
with $C_\kappa=2\omega$. In other words, Hamiltonians $\tilde H_\pm(\kappa)$ are {\it shape invariant} \cite{Gen}. The same is true for Hamiltonians $\hat H_\pm(\kappa)$, which, however, do not include variable parameter $\kappa$.

Thus our analysis of the realistic quantum mechanical system having a clear physical meaning (charged particle with spin 1/2 interacting with the constant and homogeneous magnetic field) make it possible to discover its  nice hidden symmetry, i.e., the shape invariance. It happens that this symmetry is valid for many other important QM systems like the Hydrogen atom, and causes their exact solvability \cite{Gen}.

To be shape invariant, Hamiltonian should be factorizable, i.e., to admit representation (\ref{n9}), (\ref{n11}) for $\tilde H_-(\kappa)$ with some function $W$ called {\it superpotential}. In addition, it should satisfy condition (\ref{n13}) together with the corresponding Hamiltonian $\tilde H_+(\kappa)$ which is called {\it superpartner}. If so, the related eigenvalue problem (\ref{00}) is exactly solvable, and its solutions can be found algorithmically.

The shape invariance condition can be formulated as a condition for the potential. Considering the 1d Hamiltonian $H=-\frac{\p2}{\p x^2}+V(\kappa,x) $ with a given potential $V$ dependent on $x$ and parameter $\kappa$ and representing $V(\kappa,x)$ as
\begin{equation}\la{n14}V=W_\kappa^2+W_\kappa'\end{equation}
where $W_\kappa'=\frac{\p W_\kappa}{\p x}$, and   superpotential is a solution of the Riccati equation (\ref{n14}). Then we construct a superpartner potential
\begin{equation}\la{n15}\tilde V=W_\kappa^2-W_\kappa'.\end{equation}
The corresponding stationary Schr\"odinger equation is shape invariant provided $\tilde V(\kappa, x)=V(\kappa+1)+C_\kappa$ where $C_\kappa$ is a constant. In terms of the superpotential this condition looks as follows:
\begin{equation}\la{n16} W_\kappa^2-W_\kappa'=W_\kappa^2+W_\kappa'+C_\kappa.\end{equation}

A natural question arises wether it is possible to formulate the shape invariance condition with another transformation law for potential parameters. The answer is yes, but the rule $\kappa\to\kappa+1$ can be treated as general up to redefinition of these parameters. In other words, we always can change these parameters by some functions of them in such a way that their transformations will be reduced to shifts \cite{Yuras}.

\section{Matrix superpotentials\label{MS}}
\subsection{Pron'ko-Stroganov problem\label{PS}}
The supersymmetric systems considered in the above include matrix potentials. However, when speaking about shape invariance, we deal with scalar potentials and superpotentials, refer to equations (\ref{n7}). Let us show that the concept of shape invariance can be extended to the case of matrix superpotentials.

Like in Section 2 we will start with a well defined QM system  which includes a matrix potential and  appears to be shape invariant. Namely, let us consider a neutral QM particle with non-trivial dipole momentum (e.g., neutron), interacting with the magnetic field generated by  by a
straight  line current directed along the third coordinate axis (Pron'ko-Stroganov problem \cite{Pron}) The corresponding
Schrodinger-Pauli Hamiltonian looks as follows:
\begin{equation}{\cal H}=\frac{p_1^2+p_2^2}{2m}+\lambda\frac{\sigma_1x_2-\sigma_2x_1}{\tilde r^2}
\label{ps2}\end{equation}  where $\lambda$ is the integrated coupling
constant, $\sigma_1$ and $\sigma_2$ are Pauli matrices.

The last term in (\ref{ps2}) is the Pauli interaction term
$\lambda{\sigma_i}{H_i}$ where the magnetic field ${\bf H}$ has
the following components which we write ignoring the constant multiplier included into the parameter $\lambda$:
\begin{equation}H_1\sim\frac{y}{r^2},\quad H_2~\sim-\frac{x}{r^2}, \quad
H_3=0.\label{ps2a}\end{equation}

Hamiltonian (\ref{ps2}) commutes with the third component of the total orbital momentum $J_3=x_1p_2-x_2p_1+\frac12\sigma_3$, thus the corresponding stationary Schr\"odinger equation (\ref{00}) admits solutions in separated variables. Moreover, the equation for radial functions takes the following form
\begin{equation}
\label{eq}
\hat{H}_\kappa\psi=E_\kappa\psi
\end{equation}
where $\hat{H}_\kappa$ is a Hamiltonian with a matrix potential,
$E_\kappa$ and $\psi$ are  its eigenvalue and eigenfunction
correspondingly, moreover, $\psi$ is a two-component spinor. Up to
normalization of the radial variable $\tilde r$ the Hamiltonian $\hat{H}_\kappa$
can be represented as
\begin{equation}
\label{Ham}
\hat{H}_\kappa=-\frac{\partial^2}{\partial \tilde r^2}+\kappa(\kappa-\st)
\frac{1}{\tilde r^2}+\so\frac{1} {{\tilde r}}
\end{equation}
where $\so$
and $\st$ are Pauli matrices and $\kappa$ is a natural number. In
addition, solutions of equation (\ref{eq}) must be normalizable and
vanish at $\tilde r=0$.

Hamiltonian $\hat{H}_\kappa$ can be factorized as in (\ref{n9})
where
\[
a_\kappa^-=\frac{\partial}{\partial \tilde r}+W_\kappa,\quad a_\kappa^+=-
\frac{\partial}{\partial \tilde r}+W_\kappa,\quad
c_\kappa=-\frac{1}{(2\kappa+1)^2}
\]
and $W$ is a {\it matrix
superpotential}
\begin{equation}
\label{s4}
W_\kappa=\frac{1}{2\tilde r}\st-\frac{1}{2\kappa+1}\so-\frac{2\kappa+1}{2\tilde r}.
\end{equation}
It is easily verified that the superpartner of Hamiltonian $\hat{H}_\kappa$ satisfies relation (\ref{n13}). In other words,
 equation (\ref{eq}) admits
supersymmetry with shape invariance and can be solved using the
standard technique of SSQM exposed, e.g., in survey \cite{Khare}.

\subsection{Generic matrix superpotentials\label{MP}}

Following a natural desire to find other  shape invariant matrix potentials we return to conditions (\ref{n16}) which should be satisfied by the corresponding matrix superpotentials.

Assume $W_k(x)$ be Hermitian. Then the corresponding
potential $V_k(x)$ and its superpartner $V^+_k(x)$, are Hermitian too.

The problem of classification of shape invariant
 superpotentials, i.e., $n\times n$ matrices whose elements are functions
 of $x, k$  satisfying conditions (\ref{n16}), was formulated and partially solved  in papers \cite{138} and \cite{139}. Here we present the completed classification results  for a special class of
 superpotentials being $2\times2$ matrices .

Consider superpotentials  of  the following special form
\begin{equation}
\label{SP} W_k=kA +\frac1k B+C
\end{equation}
where $C$, $B$ and $A$  are  Hermitian matrices depending
on $x$.

 Substituting (\ref{SP}) into (\ref{n16}) we obtain the following equations
 for $C$, $B$ and $A$:
\begin{equation}
 A'=\alpha( A^2+\nu I),
\label{a0}\end{equation}\begin{equation}\label{a00} C'-\frac\alpha2
\{ A,C\}+\kappa I=0,\end{equation}\begin{equation}\begin{array}{l} \{B,C\}+\lambda I=0,\quad
B^2=\omega^2I\end{array}\label{a8}
\end{equation}
where $ A'=\frac{d A}{dx},\quad \{ A,C\}= AC+C A$ is the
anticommutator of matrices $ A$ and $C$, $I$ is the unit matrix and
$\kappa, \ \lambda,\ \nu, \  \omega $ are constants. Thus the problem of classification of matrix superpotentials is reduced to solution of equations (\ref{a00})--(\ref{a8}) for unknown matrices $ A$ and $C$, $B$.
\subsection{Scalar superpotentials}
First we consider the scalar case when $A, C$ and $B$ in (\ref{SP})
are $1\times1$ "matrices". The corresponding equations
(\ref{a0})--(\ref{a8}) can be integrated rather easily, refer to \cite{139} for detailed calculations. As a result we obtain the well known list of scalar superpotentials:
\begin{gather}\begin{array}{ll}W=-\frac{\kappa}{x}+
\frac{\omega}{\kappa}&\texttt{(Coulomb)},
\\W=\lambda\kappa\tan\lambda
x+\frac\omega\kappa&\texttt{(Rosen1)},
\\W=\lambda\kappa\tanh\lambda
x+\frac\omega\kappa&\texttt{(Rosen2)},
\\W=-\lambda\kappa\coth\lambda
x+\frac\omega\kappa&\texttt{(Eckart)},
\\
W=\mu x & \texttt{(Harmonic
Oscillator)},
\\W=\mu x -\frac{\kappa}x&\texttt{(3D
Oscillator)},
\\W=\lambda\kappa\tan\lambda
x+\mu\sec{\lambda x}& \texttt{(Scarf
I)},
\\W=\lambda\kappa\tanh\lambda x+\mu\texttt{sech}
{\lambda x}& \texttt{(Scarf
2)},
\\W=\lambda\kappa\coth\lambda x +
\mu\texttt{cosech}\lambda
x&\texttt{(Generalized P\"oschl-Teller)},
\\W=\kappa-\mu\exp(-x)&
(\texttt{Morse}).\end{array}\label{coulomb}\end{gather}

Thus integrating equations
(\ref{a0})--(\ref{a8}) we recover the known list of
superpotentials (\ref{coulomb})
which generate classical additive shape
invariant potentials, in a straightforward and very simple way. The corresponding
potentials $V_\kappa$ can be found using definition
(\ref{n14}).
\subsection{Matrix superpotentials of dimension $2\times2$}
Here we consider the case when superpotentials are $x$-dependent $2\times 2$
matrices of form (\ref{SP}).

 Supposing that $A(x)$ is  diagonal (like in (\ref{s4})), it is possible to specify five inequivalent solutions of equations (\ref{n16}):
%\begin{equation}
%\label{constant} W_k=\gamma k\sz+\varphi\so+(\phi+\frac{\omega}{k})\st \\
%\label{inverse}\end{equation}\tau
\begin{equation}\label{inverse}\begin{array}{l}
W_{\nu,\mu}=\left(\left(2\mu+1\right)\st-2\nu-1\right)\frac1{2x}+
\frac{\omega}{2\nu+1}\so,  \quad \mu>-\frac12,\end{array}
\end{equation}\begin{equation}
\label{exp}\begin{array}{l} W_{\nu,\mu}= \lambda\left(-\nu+ \mu\exp(-\lambda
x)\so-\frac{\omega}{\nu}\st\right),\end{array}\end{equation}\begin{equation}
\label{tan}\begin{array}{l} W_{\nu,\mu}=\lambda\left(\nu\tan\lambda x
+\mu\sec\lambda x\st+\frac{\omega}{\nu}\so\right), \end{array}\end{equation}\begin{equation}
\label{cotanh}\begin{array}{l} W_{\nu,\mu}=\lambda\left(-\nu \coth\lambda x+
\mu\mathrm{csch}\lambda x\st-\frac{\omega}{\nu}\so\right),\quad \mu<0,\
\omega>0,\end{array}\end{equation}\begin{equation}\begin{array}{l}
W_{\nu,\mu}=\lambda\left(-\nu \tanh\lambda x+ \mu\mathrm{sech}\lambda
x\so-\frac{\omega}{\nu}\st \right)\label{tanh}
\end{array}\end{equation}
where we introduce the rescaled parameter $\nu=\frac{\kappa}{\alpha}.$
These superpotentials are defined up to translations $x\rightarrow
x+c$,  $\nu\rightarrow \nu+\gamma$, and up to unitary
transformations $W_{\nu,\mu}\to U_aW_{\nu,\mu}U_a^\dag$ where
$U_1=\so$, $U_2=\frac1{\sqrt{2}}(1\pm\ri \sd)$ and $U_3=\st$. In
particular these transformations change signs of parameters $\mu$
and $\omega$ in (\ref{exp})--(\ref{tanh}) and of $\mu+\frac12$ in
(\ref{inverse}), thus without loss of generality we can set
\begin{equation}\label{muo}\omega>0,\quad \mu>0\end{equation} in
superpotentials (\ref{exp})--(\ref{tanh}).

Notice that the transformations $\kappa\to \kappa'=\kappa+\alpha$ correspond to the
following transformations for $\nu$:
\begin{equation}\nu\to\nu'=\nu+1\label{kappa}.\end{equation}

If $\mu=0$ and $\omega=1$ then operator (\ref{inverse}) coincides
with the  superpotential for PS problem given by equation (\ref{s4}). For
for $\mu\neq0$ superpotential (\ref{inverse}) is not equivalent to
(\ref{s4}).
The other presented matrix  superpotentials were found in \cite{138} for the first time.

The corresponding
potentials $V_\nu$ can be found starting with
(\ref{inverse})--(\ref{cotanh}) and using definition
(\ref{n14}):
\begin{equation}
\hat V_\nu=\left(\mu(\mu+1)+\nu^2-
\nu(2\mu+1)\st\right)\frac1{x^2}-
\frac{\omega}{x}\so,\label{pot1}\end{equation}
\begin{equation}
\hat V_\nu=\lambda^2\left(\mu^2\exp(-2\lambda
x)-(2\nu-1)\mu\exp(-\lambda x)\so+2\omega
\st\right),\label{pot2}\end{equation}
\begin{equation}\begin{array}{l}
\hat V_\nu=\lambda^2\left((\nu(\nu-1)+\mu^2)\sec^2\lambda
x+2\omega\tan\lambda x\so\right.
\\\left. \hphantom{\hat V_\nu=}{} +\mu(2\nu-1)\sec\lambda x \tan\lambda
x\st \right),\end{array} \label{pot3}\end{equation}
\begin{equation}\begin{array}{l}
 \hat V_\nu=\lambda^2\left((\nu(\nu-1)+\mu^2)
\mathrm{csch}^2(\lambda x)+ 2\omega\coth\lambda
x\so\right.\\\left.\hphantom{\hat V_\nu=}{} +\mu(1-2\nu)\coth\lambda x\mathrm{csch}\lambda x\st
\right),\end{array}\label{pot5}\end{equation}\begin{equation}\begin{array}{l}
\hat V_\nu=\lambda^2\left((\mu^2-\nu(\nu-1))\mathrm{sech}^2\lambda
x+2\omega\tanh\lambda x\st\right.\\\left.
\hphantom{\hat V_\nu=}{} -\mu({2\nu}-1)
\mathrm{sech}\lambda x \tanh\lambda x\so
 \right).\end{array}
\label{pot4}
\end{equation}
Potentials (\ref{pot1}), (\ref{pot2}), (\ref{pot3}) (\ref{pot5}) and
(\ref{pot4})
 are generated by superpotentials (\ref{inverse}),
(\ref{exp}), (\ref{tan}), (\ref{cotanh})   and (\ref{tanh})
respectively.
All the above potentials are shape invariant and give rise to
exactly solvable problems for systems of Schr\"odinger--Pauli type.

It was proven in \cite{138} that $n\times n$ matrix superpotentials of the form (\ref{SP}) with a diagonal matrix $Q$ and $n>2$ can be reduced to direct sums of operators fixed in (\ref{inverse}) and scalar superpotentials  specified in equations (\ref{coulomb}). Thus in fact we present the complete description of superpotentials (\ref{SP})  being matrices of arbitrary dimension, provided matrix $Q$ is diagonal.

The case of non-diagonal matrices $Q$ has been examined in paper \cite{139}. The classifying equations (\ref{a0})--(\ref{a8}) have been solved for the cases of superpotentials being $2\times2$ or $3\times3$ matrices. In the first case the following list of superpotentials was obtained:
\begin{gather}\begin{split}&
W^{(1)}_\kappa=\lambda\left(\frac{}{}\kappa\left(\sigma_+\tan(\lambda
x+c)+\sigma_-\tan(\lambda x-c)\right)\frac{}{}\right.\\&
\left.+\mu\sigma_1\sqrt{\sec(\lambda x-c)\sec(\lambda
x+c)}+\frac{1}{\kappa}R\right), \\&
W^{(2)}_\kappa=\lambda\left(-\kappa(\sigma_+\coth(\lambda
x+c)+\sigma_-\coth(\lambda x-c))\frac{}{}\right.\\&
\left.+\mu\sigma_1\sqrt{\mathrm{csch}(\lambda x-c)\mathrm{csch}(\lambda
x+c)}+\frac{1}{\kappa}R\right),
\\&
W^{(3)}_\kappa=\lambda\left(-\kappa(\sigma_+\tanh(\lambda
x+c)+\sigma_-\tanh(\lambda x-c))\frac{}{}\right.\\&
\left.+\mu\sigma_1\sqrt{\mathrm{sech}(\lambda x-c)\mathrm{sech}(\lambda
x+c)}+\frac{1}{\kappa}R\right),
\\&
W^{(4)}_\kappa=\lambda\left(-\kappa(\sigma_+\tanh(\lambda
x+c)+\sigma_+\coth(\lambda x-c))\frac{}{}\right.\\&
\left.+\mu\sigma_1\sqrt{\mathrm{sech}(\lambda x+c)\mathrm{csch}(\lambda
x-c)}+\frac{1}{\kappa}R\right), \\&
W^{(5)}_\kappa=\lambda\left(-\kappa(\sigma_+\tanh(\lambda
x)+\sigma_-)+\mu\sigma_1\sqrt{\mathrm{sech}(\lambda x)\exp(-\lambda
x)}+\frac{1}{\kappa}R\right),
\\&
W^{(6)}_\kappa=\lambda\left(-\kappa(\sigma_+\coth(\lambda
x)+\sigma_-)+\mu\sigma_1\sqrt{\mathrm{csch}(\lambda x)\exp(-\lambda
x)}+\frac{1}{\kappa}R\right),
\\&
W^{(7)}_\kappa=-\kappa
\left(\frac{\sigma_+}{x+c}+\frac{\sigma_-}{x-c}
\right)+\frac{\mu\sigma_1}{\sqrt{x^2-c^2}}+ \frac{1}{\kappa}R,
\\&
W^{(8)}_\kappa=-\kappa \frac{\sigma_+}{x}
+\mu\sigma_1
\frac{1}{\sqrt{x}}+ \frac{1}{\kappa}R,
\\& W^{(9)}_\kappa= \lambda\left(-\kappa I+ \mu\exp(-\lambda
x)\so-\frac{\omega}{\kappa}\st\right).\end{split}\label{more}
\end{gather}
Here
\begin{equation}\label{s+-}\sigma_\pm=\frac12(\sigma_0\pm\sigma_3),\quad R=r_3\sigma_3+r_2\sigma_2,\end{equation}
$r_a$ are constants satisfying $r_2^2+r_3^2=\omega^2$,
$\kappa,\
\mu$,\   $\lambda$ and $c\neq0$ are arbitrary parameters.

\subsection{Matrix superpotentials of dimension $3\times3$}

In analogy with the above we can find superpotentials realized by irreducible $3\times3$ matrices \cite{139}, which are presented in the following formulae:
\begin{equation}\begin{array}{l}W=(S_1^2-1)\frac\kappa{x+c_1}+(S_2^2-1)\frac\kappa{x+c_2}+
(S_3^2-1)\frac\kappa{x}\\+ S_1\frac{\mu_1}{\sqrt{x(x+c_1)}}+
S_2\frac{\mu_2}{\sqrt{x(x+c_2)}}+\frac\omega\kappa
(2S_3^2-1),
\\
W=(S_1^2-1)\frac\kappa{x}+(S_2^2-1)\frac\kappa{x+c_1}+
S_1\frac{\mu_2}{\sqrt{x}}+
S_2\frac{\mu_1}{\sqrt{x+c_1}}+\frac\omega\kappa
(2S_3^2-1),
\\W=(S_1^2-1)\frac\kappa{x+c_1}+(S_3^2-1)\frac\kappa{x}+
S_1\frac{\mu_2}{\sqrt{x}}+
S_3\frac{\mu_1}{\sqrt{x(x+c_1)}}+\frac\omega\kappa
(2S_3^2-1),\\W=(S_1^2-1)\frac\kappa{x}
+S_1c+ S_2\frac{\mu_1}{\sqrt{x}}+\frac\omega\kappa
(2S_3^2-1),\\W=(S_1^2-1)\frac\kappa{x+c_1}+(S_2^2-1)\frac\kappa{x+c_2}+
(S_3^2-1)\frac\kappa{x}\\+ S_1\frac{\mu_1}{\sqrt{x(x+c_1)}}+
S_2\frac{\mu_2}{\sqrt{x(x+c_2)}}+
S_3\frac{\mu_3}{\sqrt{(x+c_1)(x+c_2)}},\\W=(S_1^2-1)\frac\kappa{x}+(S_2^2-1)\frac\kappa{x+c_2}+
S_1\frac{\mu_1}{\sqrt{x}}+
S_2\frac{\mu_2}{\sqrt{x+c_2}}+S_3\frac{\mu_3}{\sqrt{x(x+c_2)}},\\W=(S_1^2-1)\frac\kappa{x} +S_1c+
S_3\frac{\mu_1}{\sqrt{x}}+S_2\frac{\mu_2}{\sqrt{x}}\end{array}\end{equation} where $c, c_1, c_2,
\mu_1 $ and $\mu_2$ are integration constants, and
\begin{equation}\label{s}S_1=\left(\begin{array}{rrr}0&0&0\\0&0&-\ri\\0&\ri&0\end{array}\right)
,\quad
S_2=\left(\begin{array}{rrr}0&0&\ri\\0&0&0\\-\ri&0&0\end{array}\right),\quad
S_3=\left(\begin{array}{rrr}0&-\ri&0\\\ri&0&0\\0&0&0\end{array}\right)
\end{equation}
are matrices of spin $s=1$.

The hermiticity condition generates the following restrictions:
\begin{equation}x>0,\quad  \texttt{if} \quad \mu_1^2+\mu_2^2>0; \qquad
c_i<0\quad \texttt{if} \quad \mu_i\neq0.\label{condo}\end{equation}

Formulae (\ref{inverse})-(\ref{tanh}),  (\ref{more}) give the completed list of the certain class of matrix superpotentials. Note that they give rise to many realistic QM models described by coupled systems of Schr\"odinger equations, see the following section.

\subsection{Shape invariant QM systems with matrix potentials \la{examples}}

The discussed matrix superpotentials naturally appear in realistic QM systems. The entire collection of such system can be found in \cite{Nik6}, \cite{Nik7} and \cite{Nik4}. Here we present two examples only.

Consider the following Hamiltonian
\begin{equation}\la{Eq}
H=\frac{p^2}{2m}+\frac\lambda{2m}
\sigma_i B_i+V
\end{equation}
were
$\sigma_i$ are Pauli
matrices, ${ B_i}={ B_i}({\bf x})$ are  vector components of  magnetic field strength, $V=V({\bf x})$ is a potential and
vector $\bf x$ represents independent variables. In addition,
$\lambda$ denotes the constant of anomalous coupling which is
usually represented as $\lambda=g\mu_0$ where
 $\mu_0$ is the Bohr magneton and $g$
is the Land\'e factor.

Formula (\ref{Eq}) presents a generalization of the Pron'ko-Stroganov Hamiltonian for the case of arbitrary external field. And some Schr\"odinger equations with Hamiltonians (\ref{Eq}) appears to be shape invariant. The example is given by the following equation:
\begin{equation}H\psi\equiv(-\nabla^2+\lambda(1-2\nu)\exp(-x_2)(\sigma_1\cos
    x_1-\sigma_2\sin
    x_1)\\+\lambda^2\exp(-2x_2))\psi=\hat E\psi.
    \label{EP1}\end{equation}
    %\end{widetext}
    Here $\lambda$ is the integrated coupling constant, and independent variables are rescalled to obtain more compact formulae.

    Hamiltonian $H$ in (\ref{EP1}) admits integral of motion $Q=p_1-\frac{\sigma_3}2$. Thus it is possible to expand solutions of
(\ref{EP1}) via eigenvectors of $Q$ which look as follows:
\begin{equation}\label{psi}\psi_p=\left(\begin{array}{cc}
\exp(i(p+\frac12)x_1)\varphi(x_2)\\\exp(i(p-\frac12)x_1)
\xi(x_2)\end{array}\right)\end{equation} and satisfy the condition
$Q\psi_p=p\psi_p$.

Substituting (\ref{psi}) into (\ref{EP1}) we come to the following equation
$$
\hat H_\nu\psi\equiv\left(-\frac{\p^2}{\p x^2}+\hat V_\nu\right)
\psi=E\psi$$
 where
 \begin{equation}\label{V}\begin{array}{l}\hat V_\nu=\lambda^2\exp(-2y)-\lambda(2\nu-1)\exp(-y)\sigma_1-p\sigma_3,
\\y=x_2,\quad E=\tilde E-p^2-\frac14,\quad
\psi=\left(\begin{array}{l}\varphi\\
\xi\end{array}\right)\end{array}
\end{equation}

 Potential $\hat V_\nu$
(\ref{V}) belongs to the list of shape invariant matrix potentials
presented in the above, see equation (\ref{pot2}).
Thus equation (\ref{EP1}) can be solved exactly using tools of SUSY quantum mechanics \cite{Nik7}. Notice that this equation is also superintegrable \cite{Nik6}.

 Let us present an analogue of the PS model for particle
 of spin $1$.
 This model is both superintegrable and shape invariant.
 It is based on the following Hamiltonian
\begin{equation}{\cal H}_s=\frac{p_1^2+p_2^2}{2m}+\frac1{r}\mu_s(
{\bf n}) \label{ps21}\end{equation} where
\begin{equation}\label{mus} \mu_s( {\bf n})=\mu_1( {\bf n})=\mu (2({\bf
S}\times{\bf n})^2-1) +\lambda (2({\bf S}\cdot{\bf
n})^2-1).\end{equation} Here $\mu$ and $\lambda$ are arbitrary real
parameters,  ${\bf S}\cdot{\bf n}=S_1n_2+S_2n_1$ and ${\bf
S}\times{\bf n}=S_1n_2-S_2n_1$, $S_1$ and $S_2$ are matrices of spin
1 given by formula (\ref{s}).

It is the Hamiltonian defined by equations (4.1) and (4.2) that
generalized  the Pron'ko-Stroganov model  for the case of spin one. This Hamiltonian leads to shape invariant radial equations with matrix potential being the direct sum of a modified Coulomb potential and potential (\ref{pot1}).

\subsection{Dual shape invariance \label{dsi}}

Starting with (\ref{inverse})--(\ref{cotanh}) we found
the related potentials (\ref{pot1})--(\ref{pot5}) in a unique
fashion. But there an interesting  inverse problem: to find possible
superpotentials corresponding to given potentials. Formally speaking, this means to find all solutions of the Riccati equation (\ref{n14}) for $W$. However, such solutions depend on two arbitrary parameters ($\nu$ and the integration constant), and there is some ambiguity in choosing such of them which should be changing to generate the superpartner potential. Notice that the mentioned inverse problem is
 very interesting since it opens a way to
generate families of isospectral hamiltonians \cite{Khare}.

In the case of
 matrix superpotentials this business is even more important since in some cases there exist two superpotentials compatible
with the shape invariance condition. And both these  superpotential can be requqested to generate solutions of the related eigenvalue problem.

To find the mentioned additional superpotentials we use the invariance of potentials (\ref{pot1}), (\ref{pot3}) and
(\ref{pot5}) with respect to the simultaneous change of arbitrary parameters:
\begin{equation}
\label{change}
\mu \to \nu-\frac12,\quad \nu\to\mu+\frac12.
\end{equation}
This means that in addition
to the shape invariance w.r.t.\ shifts of $\nu$ potentials
(\ref{pot1}), (\ref{pot3}) and (\ref{pot5}) should be shape
invariant w.r.t.\ shifts of parameter $\mu$ too.

Thus,  it is possible to represent
potentials (\ref{inverse}),  (\ref{tan}) and (\ref{cotanh}) in the following alternative form
\begin{equation}\label{SS}\widetilde W_{\mu,\nu}^2-\widetilde W'_{\mu,\nu}={\hat  V}_{\mu}+c_\mu\end{equation}
where $\hat V_\mu=\hat V_\nu$, and\begin{equation}\label{SS1}\widetilde
W_{\mu,\nu}=\frac{\nu\st-\mu-1}{x}
+\frac{\omega}{2(\mu+1)}\so,\quad
c_\mu=\frac{\omega^2}{4(\mu+1)^2}\end{equation} for $\hat V_k$ given by
equation (\ref{pot1}),
\begin{equation}\label{SS2}
\widetilde W_{\mu,\nu}=\frac\lambda2\left((2\mu+1)\tan\lambda x
+(2\nu-1)\sec\lambda x\st+\frac{4\omega}{2\mu+1}\so\right)
\end{equation} for potential (\ref{pot3}), and
\begin{equation}\label{SS3}
 \widetilde W_{\mu,\nu}=\frac\lambda{2}\left(-(2\mu+1) \coth\lambda
x+ (2\nu-1)\mathrm{csch}\lambda
x\st-\frac{4\omega}{2\mu+1}\so\right)\end{equation} for potential
(\ref{pot5}).   The related
 constant $c_\mu$  is:
\begin{equation}\label{77}
c_\mu=\lambda^2\left(\pm\frac14(2\mu+1)^2+
\frac{4\omega^2}{(2\mu+1)^2}\right)\end{equation}
where the sign ``$+$" and ``$-$" corresponds to the cases (\ref{SS2}) and (\ref{SS3}) respectively.

We stress that superpartners of potentials (\ref{SS}) constructed
using superpotentials $\widetilde W_{\mu,\nu}$, i.e.,
\begin{equation}
\label{SSS} \hat V^+_{\mu} =\widetilde W_{\mu,\nu}^2+\widetilde
W'_{\mu,\nu}
\end{equation}
satisfy the shape invariance condition since
$$
\hat V^+_{\mu}=\hat V_{\mu+1}+C_\mu
$$
with $C_\mu=c_{\mu+1}-c_\mu$.

Thus  potentials  are shape
invariant w.r.t.\ shifts of two parameters, namely, $\nu$ and
$\mu$. More exactly, superpartners for  potentials (\ref{pot1}),
(\ref{pot3}) and (\ref{pot5}) can be obtained  either by shifts of
$\nu$ or by shifts of~$\mu$ while simultaneous shifts are
forbidden. We call this phenomena {\it dual shape invariance}.

Notice that the dual shape invariance makes it possible to construct the complete set of ground state solutions for all admissible values of quantum numbers enumerating these solutions, see Section 5.2.

\section{Exact solutions of shape invariant
Schr\"odinger equations\label{SE}}
\subsection{Generic approach and energy values}

An important consequence of the shape invariance is the nice possibility to construct exact solutions of the related stationary
Schr\"odinger equation. The procedure of construction of exact solutions for the case of scalar shape invariant potentials is described in various surveys, see, e.g., \cite{Khare}. Here we present this procedure for the more general case of matrix potentials.

Consider the stationary Schr\"odinger equation
\begin{equation}
\hat H_\nu\psi\equiv\left(-\frac{\p^2}{\p x^2}+\hat V_\nu\right)
\psi=E_\nu\psi\label{se2}
\end{equation}
where $\hat H_\nu=a^+_{\nu,\mu}a^-_{\nu,\mu}+c_\nu$ and $\hat V_\nu$ is a shape invariant
potential.  An algorithm for construction of
 exact solutions of supersymmetric and shape invariant Schr\"odinger equations
 includes the following steps (see, e.g., \cite{Khare}):
 \begin{itemize}
 \item To find the ground state solutions $\psi_0(\nu,\mu,x)$ which are proportional to
  square integrable solutions of the first order equation
  \begin{equation}
  \label{psi0}
  a_{\nu,\mu}^-\psi_0(\nu,\mu,x)\equiv
  \left( \frac{\p}{\p
  x}+W_{\nu,\mu}\right)\psi_0(\nu,\mu,x)=0.
  \end{equation}
  Function $\psi_0(\nu,\mu,x)$ solves equation
(\ref{se2}) with
\begin{equation}
E_\nu=E_{\nu,0}=-c_\nu\label{E0k}.
\end{equation}
\item To find a solution $\psi_1(\nu,\mu,x)$ for the first excited state which is
defined by the following relation:
\begin{equation}
\label{psi1}
\psi_1(\nu,\mu,x)=a^+_{\nu,\mu}
\psi_0(\nu+1,\mu,x)\equiv\left(- \frac{\p}{\p
x}+W_{\nu,\mu}\right)\psi_0(\nu+1,\mu,x).
\end{equation}
 Since $a_\nu^\pm$ and $\hat{H}_\nu$ satisfy the intertwining relations
  \begin{equation}\la{Inter}\hat H_\nu a_{\nu,\mu}^+=a_{\nu,\mu}^+ \hat H_{\nu+1}\end{equation}
  function (\ref{psi1})  solves
  equation (\ref{se2}) with $E_\nu=E_{\nu,1}=-c_{\nu+1}$.
  \item Solutions for the second excited state  $\psi_2(\nu,\mu,x)$ can be found acting by the first order differential operator $a^+$ on the first exited state, i.e.,
  $\psi_2(\nu,\mu,x)=a^+_{\nu,\mu}\psi_1(\nu+1,\mu,x)$,
  etc. Finally, solutions
  which correspond to $n^{th}$ exited state for any admissible natural number $n>0$
  can be represented as
\begin{equation}\label{psin}\psi_n(\nu,\mu,x)=
a_{\nu,\mu}^+a_{\nu+1,\mu}^+ \cdots
a_{\nu+n-1,\mu}^+\psi_0(\nu+n,\mu,x).
\end{equation} The corresponding eigenvalue $E_{\nu,n}$ is equal to
$-c_{\nu+n}$.
\item For systems admitting the dual shape invariance it is
necessary to repeat the steps enumerated above using alternative (or additional)
superpotentials.
\end{itemize}

All matrix potentials presented in the above generate integrable
models with Hamiltonian (\ref{se2}). However, it is necessary  to
examine their consistency, in particular,  to verify
that there exist square integrable solutions of equation
(\ref{psi0}) for the ground states.

In the following sections we find such solutions for all
superpotentials given by equations (\ref{inverse})--(\ref{cotanh})
and (\ref{SS1})--(\ref{SS3}). However, to obtain
normalizable ground state solutions it is necessary to impose
certain conditions on parameters of these superpotentials.

Let us present the energy spectra for models
(\ref{se2}) with potentials (\ref{pot1})--(\ref{pot5}) which can be found applying the presented algorithm:
\begin{equation}\label{EV0}E=
-\frac{\omega^2}{(2N+1)^2}\end{equation}
for potential (\ref{pot1}),
\begin{equation}\label{EV1}E=-\lambda^2\left(N^2+\frac{\omega^2}
{N^2}\right)\end{equation} for potentials (\ref{pot2}), (\ref{pot5}),
(\ref{pot4}), and{\samepage
\begin{equation}\label{EV2}E=\lambda^2\left(N^2-\frac{\omega^2}{N^2}\right)\end{equation}
for potentials  (\ref{pot3}).}

Here $N$ is the spectral parameter which can take the following
values
\begin{equation}\label{Nn}N=n+\nu,  \end{equation} and (or)
\begin{equation}\label{Nnm}N=n+\mu+\frac12  \end{equation}
where $n=0,1,2,\dots $ are natural numbers which can take any values
for potentials (\ref{pot1})--(\ref{pot3}). For potentials
(\ref{pot2}), (\ref{pot4}) and (\ref{pot5}) {\it with a fixed $\nu<0$}
the admissible values of $n$ are bound by the condition
$(\nu+n)^2>|\omega|$.

\subsection{Ground state solutions\label{GSS}}

 To find the ground state solutions for equations (\ref{se2}) with
potentials (\ref{pot1})--(\ref{pot5}) it is sufficient
 to solve equations (\ref{psi0}) where $W_{\nu,\mu}$ are
 superpotentials  (\ref{inverse})--(\ref{cotanh}),
 and analogous equation with superpotentials (\ref{SS1})--(\ref{SS3}).
 This can be done for all the mentioned cases, but we present here only two of them.

 The corresponding solutions should be square integrable two component functions which we denote as:
\begin{equation}\psi_0(\nu,\mu,x)= \left(\begin{array}{c}\varphi\\
\xi\end{array}\right)\label{psi00}.\end{equation}

Consider the superpotential defined by equation (\ref{inverse}).
Substituting (\ref{inverse}) and (\ref{psi00}) into (\ref{psi0}) we
obtain :
\begin{gather}\label{GS11}
\frac{\p \varphi}{\p
x}+\left(\mu-\nu\right)\frac{\varphi}{x}+\frac\omega{2\nu+1}\xi=0,
\\\label{GS12}
\frac{\p \xi}{\p
x}-\left(\mu+\nu+1\right)\frac{\xi}{x}+\frac\omega{2\nu+1}\varphi=0.
\end{gather}
Solving (\ref{GS12}) for $\varphi$, substituting the solution into
(\ref{GS11}) and making the change
\begin{equation}\label{bes}\xi=y^{\nu+1}\hat\xi(y),\ \ \ y=\frac{\omega
x}{2\nu+1}\end{equation} we obtain the equation
\begin{equation}\label{bessel}y^2\frac{\p^2 \hat\xi}{\p y^2}+
y\frac{\p \hat\xi}{\p y}-
\left(y^2+\mu^2\right)\hat\xi=0\end{equation} whose square integrable solution solution is proportional to the modified
 Bessel function:
\begin{equation}\label{GS13}\hat\xi=cK_{\mu}(y).
\end{equation}

 Substituting  (\ref{GS13}) into (\ref{bes}) and using
(\ref{GS12}) we obtain:
\begin{equation}\label{GS1}\varphi=y^{\nu+1}
K_{\mu+1}(y), \quad \xi=y^{\nu+1} K_{|\mu|}(y)
\end{equation} where $y$ is the variable defined in
(\ref{bes}), $\omega x/(2\nu+1)\geq0$.

 Functions (\ref{GS1}) are square integrable provided
parameter $\nu$ is positive and satisfies the following relation:
\begin{equation}\quad \nu-\mu>0\label{condk1}.\end{equation}

If this condition is violated, i.e.,
$\nu-\mu\leq0$
solutions  (\ref{GS1}) are
not square integrable. But since potential (\ref{pot1}) admits the
dual shape invariance, it is possible to make an alternative
factorization of equation (\ref{se2}) using superpotential
(\ref{SS1}) and search for normalizable solutions of the following
equation:
\begin{equation}\tilde a_{\mu,\nu}^-\tilde\psi_0(\mu,\nu,x)\tilde\psi_0(\mu,\nu,x)=0.\label{psi0m}\end{equation}
   where $\tilde a^-_{\mu,\nu}= \frac{\p}{\p
  x}+\widetilde W_{\mu,\nu}.$
   Indeed, solving (\ref{psi0m})
   we obtain a perfect ground state vector:
  \begin{equation}\label{SS10}\tilde\psi_0(\mu,\nu,x)=
\left(\begin{array}{l} \tilde\varphi\\
\tilde\xi\end{array}\right),\quad \tilde\varphi=y^{\mu+\frac32}
K_{|\nu|}\left(y\right), \quad\tilde\xi=
y^{\mu+\frac32}K_{|\nu-1|}\left(y\right)
\end{equation} where $y=\frac{\omega x}{2(\mu+1)}$ and  $\nu=\nu+1/2.$
The normalizability conditions for solution (\ref{SS10}) are:
\begin{equation}\nu-\mu<1,\quad \mathrm{if}\quad
\nu\geq0,\mathrm{ and}\quad
\nu+\mu>1,\quad \mathrm{if}\quad
\nu<0.\label{SS55}\end{equation}

 Analogously, considering equation (\ref{psi0}) with superpotential
(\ref{exp}) and representing its solution in the form (\ref{psi00})
with
$$\xi=y^{\frac12-\nu}\hat\xi(y),\quad
\varphi=y^{\frac12-\nu}\hat\varphi(y), \quad  y=\mu\exp(-\lambda
x)$$
 we find the following solutions: \begin{equation}\label{GS4}
\varphi=y^{\frac12-\nu}K_{|\nu|}( y),\quad
\xi=-y^{\frac12-\nu}K_{|\nu-1|}(y)\end{equation}where
$\nu=\omega/\nu+1/2$ and parameters $\omega$ and $\nu$ should
satisfy the conditions
\begin{equation}  \nu<0,\quad \nu^2>\omega.\label{th1}
\end{equation}

 Since potential (\ref{pot2}) does not admit the dual shape
 invariance, there are no other ground state solutions.

  In analogous manner we find solutions of equations (\ref{psi0}) and
  (\ref{psi0m}) for the remaining superpotentials
  (\ref{exp})--(\ref{cotanh}), refer to \cite{138} for details. Solutions which correspond to $n^{\rm th}$ energy level can be obtained applying equation
(\ref{psin}). Under certain conditions on spectral parameters all such solutions are square integrable and reduce to zero at $x=0$ \cite{138}.

\subsection{ Isospectrality\label{iso}}
Let us note that for some values of parameters $\mu$ and $\nu$
potentials (\ref{pot1})--(\ref{pot4}) are isospectral with direct
sums of known scalar potentials.

Considering potential (\ref{pot1}) and using its dual shape
invariance it is possible to show  that for  half integer $\mu$ potential
$V_\nu$ can be transformed to a direct sum of scalar Coulomb
potentials.
In analogous way we can show that potentials (\ref{pot3}) with
 half integer $\nu$ or  integer $\mu$ is
isospectral with the potential
\begin{equation}\hat V_\nu=\lambda^2\left(r(r-1)
\sec^2\lambda x+2\omega\tan\lambda x\so\right), \quad
r=\frac12\pm\mu\quad  \mathrm{or}\quad r=\nu
\label{pot33}\end{equation} which is equivalent to the direct sum of
two trigonometric Rosen--Morse potentials. Under the same conditions
for parameters $\mu$ and $\nu$ potential (\ref{pot4}) is
isospectral with the direct sum of two Eckart
potentials. Finally, potential (\ref{pot4}) is isospectral with
 direct sum of
two hyperbolic Rosen--Morse potentials.

In other words, for some special values of parameters $\mu$ and $\nu$ there exist the isospectrality relations of matrix potentials
(\ref{pot1})--(\ref{pot4}) with well known scalar potentials. However, for another values of these parameters such relations do not exist.

\section{Shape invariant systems with position dependent mass}
SE with position dependent
 mass
are requested for description of  various condensed-matter systems
such as
 semiconductors,
quantum liquids and metal clusters,
quantum dots,
etc,  etc. However, in contrast with standard QM systems, their symmetries, supersymmetries and integrals of motion   were never investigated systematically.

The systematic study of symmetries of the position dependent mass SEs started recently. In particular, the completed  group classification of such equations in two and three dimensions have been carried out in \cite{NZ},  \cite{zasada2} and \cite{155}.  In addition, it has been shown in \cite{159} that the position dependent mass SEs are exactly solvable provided they admit a six parameter invariance group or more extended Lie symmetry. Here we present the classification of all rotationally invariant systems admitting second order integrals of motion \cite{Nik11}  which appear to be shape invariant and exactly solvable also.

\subsection{Rotationally invariant systems }
We will study stationary  Schr\"odinger equations with position dependent mass, which formally coincide with (\ref{eq}), but include Hamiltonians with variable mass parameters:
\begin{equation}\la{H}\hat H=p_af({\bf x})p_a+\tilde V({\bf x}).\end{equation}
Here  $V({\bf x})$ and $f({\bf x})=\frac1{2m({\bf x})}$ are arbitrary functions associated with the effective potential and inverse effective PDM, and summation from 1 to 3 is imposed over the repeating index $a$. In addition, ${\bf x}=(x^1,x^2,x^3),$ denotes a 3d space vector.

In paper \cite{NZ} all Hamiltonians (\ref{H}) admitting  first order integrals of motion are classified. In particular, the rotationally invariant systems include the following functions
$f$ and $V$:
\begin{equation}\la{fV} f=f(x),\quad \tilde V=\tilde V(x),\quad x=\sqrt{x_1^2+x_2^2+x_3^2}.\end{equation}

In accordance with \cite{NZ} there are four Hamiltonians which are rotationally invariant and admit a more extended symmetry with respect to continuous groups of transformations. The corresponding Schr\"odinger equations admit second order integrals of motion belonging to the enveloping algebras of the Lie algebras of these transformation groups. Such Hamiltonians  are specified by the following inverse masses and  potentials:
\begin{equation}\la{fV1}f=x^2, \quad \tilde V=0,\end{equation}\begin{equation}f=(1+x^2)^2,
\quad \tilde V=-6x^2, \la{fV2} \end{equation}\begin{equation}f=(1-x^2)^2, \quad
\tilde V=-6x^2,\la{fV3} \end{equation}\begin{equation}
f=x^4, \quad  \tilde V=-6x^2.
\la{fV4}\end{equation}

All PDM systems admitting second order integrals of motion are classified  in \cite{Nik11}. There are two subclasses of such systems. One class include the systems admitting vector integrals of motion while in the  second one we have the tensor integrals. All these systems are shape invariant, and are presented in the following classification Tables 1 and 2.

In the third columns of the tables the effective radial  potentials are indicated which appear after the separation of variables. All radial potentials are scalar and shape invariant, i.e., can be expressed in the form (\ref{n14}) where the related superpotentials $W_\nu$ are enumerated  in formulae (\ref{coulomb}). The kinds of the superpotentials  are fixed in the fifth columns. Notice that the trigonometric and hyperbolic  P\"oshl-Teller potentials are equivalent to the Scraft 1 and generalized  P\"oshl-Teller potentials (presented in (\ref{coulomb})) correspondingly.

The content of the terms presented in the fourth columns is explained in the next section.

We see that there exist exactly twenty superintegrable systems invariant with respect to 3d rotations. Moreover, the majority  of them is defined up to one arbitrary parameter while there exist four systems dependent on two parameters, see Items 9 and 10 in both tables.
\begin{table}
\caption{Functions $f$ and $V$ specifying non-equivalent
Hamiltonians (\ref{H}) which admit vector
integrals of motion.}
\begin{tabular}{|c|c|c|c|c|}
\hline
No&$f$&$V$&$\begin{array}{c}\texttt{Solution}\\\texttt{approach}\end{array}$&
$\begin{array}{c}\texttt{Effective}\\\texttt{potentials}\end{array}$\\
\hline
&&&&\\
1.&$x$&$\alpha x$&$\begin{array}{c}
\texttt{direct or}\\\texttt{two-step}\end{array}$&$\bea{c}\texttt{3d oscillator} \\\texttt{or Coulomb}\eea$\\
&&&&\\
2.&$x^4$&$\alpha x$&$\begin{array}{c}
\texttt{direct or}\\\texttt{two-step}\end{array}$&$\bea{c}\texttt{Coulomb or}\\\texttt{3d oscillator}\eea$\\
&&&&\\
3.&$x(x-1)^2$&$\displaystyle\frac{\alpha x}{(x+1)^2}$&$\displaystyle\begin{array}{c}
\texttt{direct or}\\\texttt{two-step}\end{array}$&$\bea{c} \texttt{Eckart or}\\\bea{c}\texttt{hyperbolic}\\\texttt{P\"oschl-Teller}\eea
\eea$\\
&&&&\\
4.&$x(x+1)^2$&$\displaystyle\frac{\alpha x}{(x-1)^2}$&$\begin{array}{c}
\texttt{direct or}\\\texttt{two-step}\end{array}$&$\bea{c}\texttt{Eckart or }\\\texttt{trigonometric}\\
\texttt{P\"oschl-Teller}\eea$\\&&&&\\
5.&$(1+x^2)^2$&$\displaystyle\frac{\alpha (1-x^2)}{x}$&\texttt direct&$\bea{l}\texttt{trigonometric}\\
\texttt{Rosen-Morse}\eea$
\\
6.&$(1-x^2)^2$&$\displaystyle\frac{\alpha (1+x^2)}{x}$&\texttt {direct}&$\bea{c}\texttt{ Eckart}
\eea$\\&&&&\\
7.&$\displaystyle\frac{x}{ x+1}$&$\displaystyle\frac{\alpha x}{
x+1}$ &\texttt {two-step}&$\bea{c}
\texttt{Coulomb }\eea$\\&&&&\\
8. &$\displaystyle\frac{x}{x-1}$&$\displaystyle\frac{\alpha x}{ x-1}$&\texttt
{two-step}&\texttt{Coulomb} \\
&&&&\\
9.&$\bea{l}\displaystyle\frac{(x^2-1)^2x}{x^2-2\nu x+1}\eea$&$\displaystyle\frac{\alpha x}{x^2-2\nu x+1}$&\texttt {two-step}&
$\bea{l}\texttt { Eccart}
\eea$\\
&&&&\\
10.&$\displaystyle\frac{(x^2+1)^2x}{x^2-2\nu x-1}$& $\displaystyle\frac{\alpha
x}{x^2-2\nu x-1}$&\texttt {two-step}&$\bea{l}\texttt{trigonometric}\\
\texttt{Rosen-Morse}\eea$
\\
\hline
\end{tabular}
\end{table}

\vspace{2mm}

\begin{table}
\caption{Functions $f$ and $V$ specifying non-equivalent
Hamiltonians (\ref{H}) which admit tensor
integrals of motion.}
\begin{tabular}{|c|c|c|c|c|}
\hline\noalign{\smallskip}
No&$f$&$V$&$\begin{array}{c}\texttt{Solution}\\\texttt{approach}\end{array}$&
$\bea{c}\texttt{Effective}\\\texttt{radial}\\\texttt{potential}\eea$\\
\hline
&&&&\\
1.&$\displaystyle\frac{1}{x^2}$&$\displaystyle\frac\alpha{x^2}$&$\begin{array}{c}
\texttt{direct}\\\texttt{or two-step}\end{array}$&$\bea{c}\texttt{Coulomb or}
\\\texttt{3d oscillator}\eea$\\&&&&\\
2.&$x^4$&$\displaystyle-\frac\alpha{x^2}$&$
\begin{array}{c}
\texttt{direct}\\\texttt{or two-step}\end{array}$&$\bea{c}\texttt{3d oscillator}
\\\texttt{or Coulomb}\eea$\\&&&&\\
3.&$(x^2-1)^2$&$\displaystyle\frac{\alpha x^2}{(x^2+1)^2}$&$\begin{array}{c}
\texttt{direct}\\\texttt{or two-step}\end{array}$&$\bea{c}\texttt{Eckart or}\\\texttt{hyperbolic}\\
\texttt{P\"oschl-Teller}\eea$\\&&&&\\
4.&$(x^2+1)^2$&$\displaystyle\frac{\alpha x^2}{(x^2-1)^2}$&$\begin{array}{c}
\texttt{direct}\\\texttt{or two-step}\end{array}$&$\bea{c}\texttt{Eckart or}\\\texttt{trigonometric}\\
\texttt{P\"oschl-Teller}\eea$\\
5.&$\displaystyle\frac{(x^4-1)^2}{x^2}$&$\displaystyle\frac{\alpha(x^4+1)}{x^2}$&\texttt{direct}&$\bea{c}\texttt{ Eckart}\\\eea$\\
&&&&\\
6.&$\displaystyle\frac{(x^4+1)^2}{x^2}$&$\displaystyle\frac{\alpha(x^4-1)}{x^2}$&\texttt{direct}&$\bea{c}
\texttt{trigonometric}\\\texttt{Rosen-Morse}\eea$\\&&&&\\
7.&$\displaystyle\frac1{x^2+1}$&$\displaystyle\frac{\alpha }{x^2+1}$&\texttt{two-step}&$\bea{c}\texttt{3d oscillator}\eea$\\&&&&\\
8.&$\displaystyle\frac1{x^2-1}$&$\displaystyle\frac{\alpha }{x^2-1}$&\texttt{two-step}&$\bea{c}\texttt{3d oscillator}\eea$\\&&&&\\
9.&$\displaystyle\frac{(x^4-1)^2}{x^4-2\nu x^2+1}$&$\displaystyle\frac{\alpha x^2}{x^4-2\nu x^2+1}$&\texttt{two-step}&$\bea{c}\texttt{ Eckart}\eea$\\&&&&\\
10.&$\displaystyle\frac{(x^4+1)^2}{x^4-2\nu x^2-1}$&
$\displaystyle\frac{\alpha x^2}{x^4-2\nu x^2-1}$&\texttt{two-step}&$\bea{c}
\texttt{trigonometric}\\\texttt{Rosen-Morse}\eea$\\&&&&
\\
\hline \end{tabular}\end{table}

\vspace{2mm}

%\newpage

\subsection{Two strategies in construction of exact solutions}

Let us consider  equations  (\ref{eq}) where $H$ are hamiltonians (\ref{H}) whose mass and potential terms are specified in the presented tables. We will search for square integrable solutions of these systems vanishing at $x=0$.

 First let us transform (\ref{eq}) to the following equivalent form
 \begin{equation}\la{se1}
   \tilde H \Psi=E \Psi,
\end{equation}
where
  \begin{equation}\la{Hef} \tilde H=\sqrt{f}H \frac{1}{\sqrt{f}}=fp^2+V,
  \quad \Psi=\sqrt{f}\psi. \end{equation}

  Then, introducing spherical variables and expanding solutions via spherical functions $Y^l_m$
 \begin{equation}\la{rv}\Psi=\frac1x\sum_{l,m}\phi_{lm}(x)Y^l_m\end{equation}
we obtain the following equation for radial functions:
\begin{equation}\la{re}-f\frac{\p^2\phi_{lm}}{\p x^2}+\left(\frac{fl(l+1)}{x^2}+V \right)\phi_{lm}=E\phi_{lm}.\end{equation}

 Let us present two
possible ways to solve equation (\ref{re}). They can be treated as particular cases of Liouville transformation (refer to \cite{olver} for definitions) and include commonly known steps. But it is necessary to fix them as concrete algorithms to obtain shape invariant potentials presented in the tables.

The first way (which we
call direct) includes consequent changes of independent and
dependent variables: \begin{equation} \phi_{lm}\to \Phi_{lm}=
f^{\frac14}\phi_{lm}, \ \frac{\p}{\p x}\to f^{\frac14} \frac{\p}{\p
x} f^{-\frac14}= \frac{\p}{\p
x}+\frac{f'}{4f}\la{change1}\end{equation} and then \begin{equation} \la{change3}
x\to y(x), \end{equation} where $y$ solves the equation $\frac{\p
y}{\p x}=\frac1{\sqrt{f}}$.  As a result equation (\ref{rv}) will be
reduced to a more customary form
\begin{equation}\la{re2}-\frac{\p^2\Phi_{lm}}{\p y^2}+\tilde V  \Phi_{lm}=
E\Phi_{lm}\end{equation} where $\tilde V$ is an effective potential
\begin{equation}\la{po}\tilde V=V+f\left({\frac{l(l+1)}{x^2}} -\left(\frac{
f'}{4 f}\right)^2-\left(\frac{ f'}{4 f}\right)'\right),\quad
x=x(y).\end{equation}

  Equations (\ref{se1}), (\ref{Hef})
with functions $f$ and $V$ specified in Items 1--6 of both Tables 1
and 2 can be effective solved using the presented reduction to
radial equation (\ref{re2}). All the corresponding potentials (\ref{po}) appears to be shape invariant, and just these potentials are indicated in the fifth columns of the tables. The related  equations
(\ref{re2}) are shape invariant too and can be solved using
the SUSY routine.

However, if we apply the direct approach to  the remaining systems (indicated in Items 7 -- 10 of both  tables), we come to equations  (\ref{re2}) which are not shape invariant and are hardly solvable, if at all.  To solve these systems we need a more sophisticated  procedure which we call the two-step approach. To apply it we multiply (\ref{re}) by $\alpha V^{-1}$
and obtain the following equation:
\begin{equation}\la{re3}-\tilde f\frac{\p^2\phi_{lm}}{\p x^2}+\left(\frac{\tilde f l(l+1)}{x^2}+\tilde V \right)\phi_{lm}={\cal E}\phi_{lm}\end{equation}
where $\tilde f=\frac{\alpha f}{ V},\ \tilde V =-\frac{\alpha E}V$ and $ {\cal E}=-\alpha$ .
Then treating $\cal E$ as an  eigenvalue and solving equation (\ref{re3}) we can find $\alpha$ as a function of $E$, which defines admissible energy values at least implicitly. To do it it is convenient to make changes (\ref{change1}) and (\ref{change3}) where $f\to \tilde f$.

The presented trick with a formal changing the roles of constants $\alpha$ and $E$ is well known.  Our
point is that {\it any of the presented superintegrable systems can
be effective solved using either the direct approach presented in
equations (\ref{Hef})--(\ref{po}), or the two-step approach}. Moreover, some of the presented systems can be solved using both the direct and two-step approaches, as indicated in the fourth columns of Table 1 and 2. In all cases we obtain shape
invariant effective potentials and can use tools of SUSY quantum
mechanics.
\subsection{System including two parameters}

Let us consider the systems specified in Item 10 of Table 2. The corresponding Hamiltonian (\ref{Hef}) and radial equation (\ref{re}) have the following form:
$$ H=\frac{(x^4+1)^2}{x^4-2\nu x^2-1}p^2 +\frac{\alpha x^2}{x^4-2\nu x^2-1}$$
and
\begin{equation}\la{re8}\left(-
\frac{(x^4+1)^2}{x^4-2\nu x^2-1}\left(\frac{\p^2}{\p x^2}-\frac{l(l+1)}{x^2}\right)+
\frac{\alpha x^2}{x^4-2\nu x^2-1}\right)\phi_{lm}= E\phi_{lm}.\end{equation}

Multiplying (\ref{re8}) from the left by $\frac{x^4-2\nu x^2-1}{x^2}$ we come to the following equation:
\begin{equation}\la{re81}\left(-
\frac{(x^4+1)^2}{x^2}\left(\frac{\p^2}{\p
x^2}-\frac{l(l+1)}{x^2}\right)+\frac{\tilde\alpha(x^4-1)}{x^2}\right)
\phi_{lm}= {\cal E}\phi_{lm}\end{equation} where
\begin{equation}\la{ev5}\tilde\alpha=-E \quad \texttt{and} \quad {\cal
E}=-\alpha-2\nu E.\end{equation}

Notice that equation (\ref{re81}) with $\tilde \alpha\to \alpha$ and
${\cal E}\to E$ is needed also to find eigenvectors of the
Hamiltonian whose mass and potential terms are specified in Item 6
of Table 2.

Making transformations (\ref{change3}) and (\ref{change1}) with $y=\frac12\arctan(x^2)$ and
\\$f=\frac{(x^4+1)^2}{x^2}$ we reduce
equation (\ref{re81}) to the following form:
\begin{equation}-\frac{\p^2\Phi_{lm}}{\p y^2} +
\left(\mu(\mu-4)\csc^2(4y)+2\tilde\alpha \cot(4y)\right)\Phi_{lm}=
\tilde{\cal E}\Phi_{lm}\la{re9}\end{equation}
where
\begin{equation}\la{EEE}\tilde {\cal E}={\cal E}+4,\quad \mu=2l+3.\end{equation}

Thus we come to equation with a shape invariant (Rosen-Morse I) potential. It is consistent provided parameters $\tilde \alpha$ and $\mu$ are positive. Solving this equation usring the standard tools ou SUSY QM we easy find its eigenfunctions and eigenvalues; the corresponding eigenvalues for equation (\ref{re8}) are given by the following formula \cite{Nik11}\begin{equation}\la{ev10}E_n=(2l+3+4n)^2\left(\nu-\sqrt{\nu^2+1
+\frac{\alpha-4}{(2l+3+4n)^2}}\ \ \right) \end{equation}
where both $n$ and $l$ are integers.

\section{Discussion\label{disscus}}

To construct QM systems with extended SUSY  we essentially use discrete symmetries, i,e, reflections and rotations to the fixed angles.

The idea itself to apply reflections to construct $N=2$ SUSY was proposed in paper \cite{Gena}. Then it was applied to generate extended supersymmetries \cite{tv}, \cite{Nik1}, \cite{Nied1}, moreover, in the latter paper the discrete rotations were applied also. In addition, using these discrete symmetries, it is possible to make a reduction of SUSY algebras as it was shown in paper \cite{bek5} and some others.

We start our discussion from these old results in order to  stress that SUSY has strong roots in quantum mechanics since a lot of important  QM models do be supersymmetric. Moreover, even the simplest SUSY model, i.e., the charged particle interacting with the uniform magnetic field in fact admits the extended supersymmetry with four supercharges \cite{Nied1}.

But the main content of the present survey are modern trends in SUSY quantum mechanics. They are the matrix formulation of the shape invariance which is requested for description of QM particles with spin interacting with external fields, and supersymmetries of Schr\"odinger equations with position dependent masses. We believe that the presented results can be treated as a challenge to generalize various branches of SUSY to the case of matrix superpotentials and position dependent masses. And it is nice that some elements of such generalizations can be already recognized in literature, see, e.g., \cite{tantan,sokol1,sokol2,AS1,AS2,IO}.

%\begin{figure}
  %\includegraphics{figure1.pdf}
%\end{figure}

%\begin{figure*}
  %\includegraphics[width=0.5\textttwidth]{figure1.pdf}
%\end{figure*}

%\begin{table}
%\begin{tabular}{lll}
%\hline\noalign{\smallskip}
%\end{tabular}
%\end{table}

%\begin{acknowledgements}
%If you'd like to thank anyone, place your comments here.
%\end{acknowledgements}



\end{document}